\documentclass[%groupedaddress,%nofootinbib,aps
twocolumn,prb]{revtex4-1}
\usepackage{amsmath}
\usepackage{epsfig,color}
\usepackage{graphicx}
\usepackage{dcolumn}
\usepackage{bm}
\usepackage{multirow}
\usepackage{bm}
\usepackage{enumerate}
\usepackage{graphicx}
\usepackage{verbatim}
\usepackage{hyperref}
\usepackage[utf8]{inputenc}
\usepackage[english]{babel}
\usepackage{amssymb,color}
\usepackage{amsfonts}
\usepackage{graphicx}
\usepackage{dcolumn}
\usepackage{float}
\usepackage{bm}
\def\prb{Phys. Rev. B}

\def\be{\begin{equation}}
\def\ee{\end{equation}}
\def\ba{\begin{eqnarray}}
\def\ea{\end{eqnarray}}

\begin{document}

\title{Superconductivity in Engineered Two-Dimensional Electron Gases}

\author{Andrey V. Chubukov}
\affiliation{School of Physics and Astronomy, University of Minnesota, Minneapolis. MN 55455}
\author{Steven A. Kivelson}
\affiliation{Department of Physics, Stanford University, Stanford, CA 94305}

\begin{abstract}
 We consider Kohn-Luttinger
 mechanism for superconductivity in a two-dimensional electron gas  confined to a narrow well between two metallic planes
 with two occupied subbands with Fermi momenta $k_{FL} > k_{FS}$.
 On the basis of a perturbative analysis, we conclude that
 non-s-wave superconductivity emerges even
 when the bands are parabolic.
 We analyze the conditions that maximize $T_c$ as a function of
the  distance to the metallic planes, the ratio $k_{FL}/k_{FS}$, and $r_s$, which measures the strength of Coulomb correlations.
The largest attraction is in p-wave and d-wave channels,
 of which p-wave is typically the strongest.
 For $r_s = O(1)$ we estimate that the dimensionless coupling $\lambda \approx 10^{-1}$, but it likely continues increasing   for larger $r_s$ (where we lose theoretical control).
   \end{abstract}
\maketitle

\section{Introduction}
\label{intro}

The subject of this paper is superconductivity in a two-dimensional electron gas (2DEG) due to screened Coulomb interaction between electrons, $\tilde V^{scr} (r)$. The Coulomb interaction is
repulsive, hence,  at a first glance,  it  cannot mediate the pairing between fermions. However, it has long been known that at large distances a screened Coulomb interaction undergoes Friedel oscillations: $\tilde V^{scr} (r) \propto \cos (2k_F r)/r^2$.
 These oscillations  are generated by  screening processes with momentum transfer near $2k_F$ and appear on top of regular screening from Coulomb to Yukawa potential.
 At large distances, the oscillating component of $V^{scr} (r)$ exceeds the Yukawa term, hence $V^{scr} (r)$ gets %occasionally
 locally overscreened, becoming
 negative in certain ranges of $r$.
  Because  different angular momentum components $ V^{scr} (\ell)$ are
convolutions of  $\tilde V^{scr} (r)$ with $r-$dependent form-factors,
$V^{scr} (\ell)$  and the dimensionless $\lambda (\ell) = N_F V^{scr} (\ell)$ are  negative, i.e., attractive, for certain values of $\ell$. (Here $N_F$ is the density of states at the Fermi level.)
Then, within the BCS approximation,
\be
k_B T_c = E_F A \exp[-1/|\lambda|]
\label{Tc}
\ee
where $E_F$ is the Fermi energy, $A$ is a number of order one, and $\lambda <0$ is the value of $\lambda (\ell)$ in the most strongly attractive pairing channel.

 This scenario
 was proposed
  in 1965 by Kohn-Luttinger (KL), and is known as the KL pairing mechanism~\cite{KL}.  In 3D, KL
  demonstrated
  that $\lambda (\ell)$ with large odd values of $l$ are
     attractive.
     %AC Suggest to keep the phrase below
     The attraction  can be traced to non-analyticity of the  fermionic polarization for momenta approaching $2k_F$ from below.
  Subsequent works found~\cite{fay,kagan1} that
  when the
   interaction
   is approximated by an on-site Hubbard interaction, all $\lambda (\ell)$ with $\ell >0$ are attractive at order $U^2$, and the
   most attractive
   $\lambda$ is in the p-wave channel.

 For a rotationally invariant 2DEG
 the situation is more complex;
 in a one-band system
 KL pairing
  does not
 occur  to order $U^2$.
 The reason is that the polarization bubble made out of free fermions
 is momentum independent
 for all $|\vec k| \leq 2k_F$.
 KL-type superconductivity still emerges, but one either have to go to third order~\cite{third_order} in $U$,
 or put the system on a lattice and re-calculate $U^2$ terms~\cite{lattice,andrey,hlubina,rks,alexandrof,allofus}. In the latter case, angular momentum is no longer a good quantum number, and the pairing interaction
 has symmetry labels corresponding to only a finite number of discrete irreducible representations consistent with  lattice symmetry.
 For example, for a tetragonal crystal, there
  are four one-dimensional irreducible representations, each of which includes an infinite number of harmonics.  For example,
   $B_{1g}$ representation
    includes ``harmonics'' of the form
   $\cos{2\phi}$
   (i.e. $\ell =2$), but also $\cos {6\phi}$ (i.e. $\ell = 6$) etc.
 For small density of fermions, within each irreducible representation
 one can still identify pairing states with an angular momentum index  according to the  dominant harmonic,
 but as density increases the
 mixing of harmonics also increases, and the pairing problem becomes progressively more complex, even at small $U$.
 While we know of no general theorems for the lattice problem, specific calculations show that one always does obtain an attraction in some irreducible representation channels~\cite{lattice,andrey,hlubina,rks,alexandrof,allofus}. However, the dimensionless coupling always turns out to be quite small
 (for no apparent reason).
 Longer range repulsions, such as a nearest-neighbor $V$, generally  reduce the pairing scale still further (and result in still more complicated pair wave functions), although they do not entirely kill the KL pairing.\cite{allofus}

 In this communication we consider the extension of 2D KL problem to the case when the rotational symmetry of low-energy states is preserved, but  there are two Fermi surfaces of different radii, and the dominant pairing interaction is the (repulsive) density-density interaction between fermions on the two Fermi surfaces.\cite{sriandme}  This case has some relevance to  Fe-based superconductors~\cite{my_review}, but our primary experimental reference will be engineered
 two-band 2DEG's with unequal Fermi momenta $k_{FL} > k_{FS}$,  confined to a narrow well (of width $w$) between two metallic ground-planes which screen the long-range part of the Coulomb interactions between electrons.   We
 show that in this situation there is an attraction in non-s-wave channels already at the leading (second) order in the dimensionless interaction parameter $r_s$, and the largest $\lambda (\ell)$ are for $\ell =1$ (p-wave) and $\ell =2$ (d-wave).
 considerately in the formal limit $r_s \to \infty$, although

 Our chief purpose here is practical -- to determine the optimal value of parameters (geometry, density, etc.) to maximize the superconducting $T_c$.
   Our approach is to situate the ``target'' 2DEG in an active electronic environment, and then integrate out everything in the environment to generate an effective interaction in terms of which we compute the appropriate dimensionless Fermi surface average of the effective interaction, $\lambda_\ell$, in various pairing channels.  The environment also  affects the prefactor $A$
     in Eq. (1).
      On general grounds we expect that $A\approx 1$ since even the effective interactions here are not greatly retarded -- this is one of many ways in which the physics here differs from that in the conventional electron-phonon mechanism, where $A \sim \hbar\omega_
      D/E_F \ll 1$ is parametrically small.  (Here  $\omega_
      D$ is a characteristic phonon frequency.)  However, even in the absence of retardation, by bad luck, it may well turn out that $A$ is considerably smaller than 1; for instance, in the 2d negative $U$ Hubbard model, where $T_c$ is maximal when $\rho(E_F) |U| \sim 1$, the value of $T_c$ extracted from quantum Monte Carlo calculations\cite{scalettar} corresponds to $A \approx10^{-1}$.  The more accurate weak-coupling calculation of $T_c$  due to KL effect in 3D also yields a rather small $A$ (Ref. \cite{kagan,kagan2}).
Since we do not make any pretext of actually computing the value of $A$, our goal should be taken to be making a plausible  estimate of the log of $T_c$:
\be
\log[E_F/k_BT_c]  \sim 1/|\lambda| \ .
\label{new_1}
\ee
Moreover, since $\lambda$ is generally an increasing function of $r_s$, even though the justification for our approach is fundamentally perturbative, we have attempted to extrapolate our results to larger $r_s$.
To do this we simply take the derived expression for $\lambda$ and evaluate it for $r_s \gg 1$ (where the expression simplifies);  we hope this gives suggestive evidence of the trends at larger $r_s$ although, naturally, for large enough $r_s$ we expect a Wigner crystal, not a superconductor.

The outline of the paper is the following.  In the next section we present the model and briefly summarize the results.  In Sec.
 \ref{method} we integrate out the environment and obtain the effective action for low-energy degrees of freedom. In Sec. \ref{pairing} we obtain pairing interactions in p-wave and d-wave channels as functions of system parameters.
  We summarize our conclusions in Sec. \ref{conclusions}.

\section{Two-Component 2DEG}
\label{2component}

\begin{figure}
  \includegraphics[width=\linewidth]{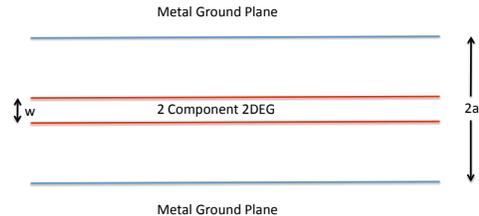}
  \caption{Schematic picture of the geometry of the two-component 2DEG considered in this paper.}
  \label{device}
\end{figure}

Because the longer range parts of the Coulomb interaction generally tend to oppose pairing,\cite{alexandrof,allofus}, we focus on a situation where these interactions are absent.
As shown in Fig. \ref{device}, we consider a 2DEG confined to a narrow well (of width $w$) between two metallic ground-planes which screen the
  interactions between electrons.
   We will assume that $w$ is small compared to the distance to the ground-plane, $w\ll a$, as this makes our analysis simpler, although this is not necessary -- all that is necessary is that the separation between the 2DEG and the metallic planes is large enough that electron tunneling is negligible.  We will also imagine we are in a range of $w$ and electron density $n$ such that there are two subbands occupied, with Fermi momenta $k_{FL}\geq k_{FS}$,
where the electron density is $n= [k_{FL}^2+k_{FS}^2]/(2\pi)$.

We will express the results in terms of a single dimensional parameter, $E_F$ (corresponding to the larger Fermi surface), and 3 dimensionless parameters,
  the dimensionless distance to the ground planes, $\alpha$, the ratio of the two Fermi momenta, $\eta$, and a conventional measure of the strength of correlations, $r_s$:
\ba
&&E_F = \hbar^2 k_{FL}^2/2m^\star\ , \ \ \alpha \equiv 2k_{FL}\  a \ , \\
&&\eta \equiv k_{FS}/k_{FL} \leq 1 \ , \ \ r_s \equiv \sqrt{2}/k_{FL}a_B^\star %\ \ {\rm with} \ \ a_B^\star \equiv \hbar^2 \kappa/e^2 m^\star \ ,
\nonumber
\ea
with $a_B^\star \equiv \hbar^2 \kappa/e^2 m^\star$ where $\kappa$ is the dielectric constant and $m^\star$ is the effective mass.  Note that $E_F$ appears only to set the energy scale in Eq. \ref{Tc}: the dimensionless factors $A$ and $\lambda$ are functions  of $\alpha$, $\eta$, and $r_s$.

On the basis of the analysis detailed  in Sec. \ref{pairing}, we have obtained the approximate integral expressions for $\lambda (l)$ as functions of $\alpha,\ \eta$, and $r_s$,  given in Eqs. \ref{ch_4} and \ref{lambda}.  These expressions are shown for various values of the parameters in Figs. \ref{fig:1na} and \ref{fig:1nb}.  Generally, we find that the dominant pairing is in the p-wave ($\ell=1$) channel.  $|\lambda (l=1)|$ is largest when $\eta \approx 1/2$, and it reaches  $\lambda (1) \approx -0.15$ when  $r_s \approx 3$. If we extrapolate the expressions to larger values of $r_s$ (which is beyond their range of demonstrated validity), $|\lambda (l=1)|$ keeps increasing and eventually can become quite large  at large $r_s$
(see Fig. \ref{fig:2} and Eq. (\ref{new_1})).

\section{General considerations}
\label{method}
To develop the formalism, consider the problem in which the electrons in the system of interest interact with the  electrons in the ``environment'' via a density-density interaction, i.e.
\be
H = H_0 + H_{env} + \int d\vec r d\vec r^\prime\ \hat \rho(\vec r)\ \tilde v_{int}(\vec r-\vec r^\prime)\  \hat\rho_{env}(\vec r^\prime)
\ee
where $H_0$ is the bare Hamiltonian for the problem at hand, $\hat \rho$ is the electron density operator, $H_{env}$ is the Hamiltonian of the active environmental degrees of freedom, and $\hat \rho_{env}$ is the corresponding charge density  operator.  Notice that here we assume that there are no important  processes that scatter an electron into or out of the ``environment;'' including such terms would change the structure of the theory somewhat.

Now integrate out the environment to obtain an effective action only for system of interest
\ba
&&S^{eff} =  S_1 -\delta\mu  \int d\tau d\vec r \ \rho(\vec r,\tau)  \\
&&\ \ \   - \frac 1 2 \int d\vec r d\vec r^\prime d\tau d\tau^\prime\  \rho(\vec r,\tau)\  \tilde D(\vec r-\vec r^\prime,\tau-\tau^\prime)\  \rho(\vec r^\prime,\tau^\prime) + \ldots \nonumber
\ea
where $S_1$ is the action of system 1 by itself (corresponding to $H_1$),
\be
\delta \mu = - \int d\vec r \ \tilde v_{int}(\vec r) \langle\  \hat \rho_{env}(\vec r, \tau)\rangle,
\ee
\be
\tilde D(\vec r,\tau) = \int d\vec r_1 d\vec r_2 \ v_{int}(\vec r-\vec r_1)\ \tilde  \chi_{env}(\vec r_1 -\vec r_2,\tau)\ v_{int}(\vec r_2),
\nonumber
\ee
with
\be
\tilde \chi_{env}(\vec r,\tau) \equiv \langle T_{\tau}\big[\hat  \rho_{env}(\vec r_1,\tau)\hat  \rho_{env}(\vec r_2,0)\big]\rangle ,
\ee
and $\ldots$ signifies higher order terms in powers of $\rho$.  Note that in Fourier transform
\be
 D(\vec k,\nu) = | v_{int}(\vec k)|^2\   \chi_{env}(\vec k,\nu).
\ee

At this stage, this set of formal manipulations could just as well have been used to describe the usual mechanism for generating an effective attraction from the electron-phonon interaction.  In that case, however, the frequency dependence of $D$ plays an essential role in the physics of superconductivity -- the effective interactions are highly retarded.  In the present case, we are imagining we are integrating out other electronic degrees of freedom, which are therefore not parametrically slower than those in the system of interest.  Indeed, since we will be focussing on a Fermi surface instability in the system of interest, in general the electronic modes we are interested in will be slow compared to the relevant environmental density fluctuations.  We therefore assume that the relevant physics involves slow density fluctuations of the system of interest, which allows us to approximately replace  the effective action by an effective Hamiltonian:
\be
H^{eff} = K + \frac 1 2 \int d\vec r d\vec r^\prime\ \hat \rho(\vec r) \tilde V^{eff}(\vec r-\vec r^\prime) \hat \rho(\vec r^\prime)
\label{Heff}
\ee
where $K$ is the ``kinetic energy,'' i.e. the band structure of the interesting electrons (including a renormalized chemical potential), and
\be
\tilde V^{eff}(\vec r) = \tilde V_{0}(\vec r) - \int d\tau \tilde D(\vec r,\tau).
\ee
 where $V_0$ is the original unrenormalized interaction.
 Equivalently,
\be
 V^{eff}(\vec k) = V_{0}(\vec k) -  D(\vec k,0).
\ee
$H^{eff}$ in Eq. (\ref{Heff}) is still an interacting problem.  To estimate the resulting value of $T_c$, we therefore solve it using BCS
 theory.  The result is an expression for $T_c$ of the form of Eq.
 (\ref{new_1}),
 with explicit expression for $\lambda$,
  as we will discuss
 below.  To the extent that the effective interactions are somewhat retarded, this set of approximations could lead to an overestimate of
  $Tc$,
   but not one that we imagine will qualitatively change our conclusions.

\subsection{Effect of placing the 2DEG in a capacitor}

If, as illustrated in Fig. \ref{device}, we place the 2DEG in a capacitor, so that there is a ground plane a distance $a$ above and below the 2DEG, this replaces the Coulomb interaction (whose Fourier transform is
$ V_0 ({\vec k}) =  2\pi e^2/|\vec k|$)
with an effective interaction which we can compute using method of images to be
\be
 V^{scr}(\vec r)= e^2 \sum_n \left[ \frac {(-1)^n}{\sqrt{|\vec r|^2 + (2na)^2}}\right]
\ee
where $\vec r$ is a $2D$ vector and $2a$ is the distance between the two capacitor plates.  The  Fourier transform of this is
\be
 V^{scr}(\vec k) = \frac {2\pi e^2}{|\vec k|} \ \tanh[|\vec k|a].
\label{ch_1}
\ee
Note that
\be
V^{scr} \sim \frac {2\pi e^2} k
\left\{
\begin{array}{ccc}
ka\left[ 1- \frac {(ka)^2} 3 \right ] & {\rm for}  &  ka \ll 1 \\
1  &  {\rm for}   &     ka \gg 1
\end{array}
\right .
\ee
and that
\be
\frac {d
V^{scr}}{dk} =- \frac {2\pi e^2}{k^2} \left[ 1-\frac{ka}{\cosh ^{2}(ka)}\right] < 0
\ee
{\it i.e.} $
V^{scr}$ is a monotonically decreasing function of $k$.  As a function of $r$, $
V^{scr} \sim e^2/r $ for $r \ll a$ while for $r
\gg a$,  $
 V^{scr} \sim  (2e^2/a) K_0(\pi r/2a) \sim
 (2e^2/\sqrt{ra}) \exp[-\pi r/2a]$.  It is  short-ranged, but still uniformly repulsive.

\section{The pairing interaction}
\label{pairing}

We now consider the situation
  in which the 2DEG in the capacitor is, itself, a two-component system with a larger and a smaller Fermi pockets.
The screened Coulomb interaction from Eq. (\ref{ch_1}) gives an effective density-density interaction for fermions near each of these two pockets.

As a warm-up, let us momentarily treat $
 V^{scr}(\vec q)$ as the pairing interaction for fermions within a given pocket, between Fermi momenta ${\bf k}$ and ${\bf p}$, $|{\bf k}|=|{\bf p}| = k_{F}$ i.e., associate
 $V^{scr}({\vec k}-{\vec p})$ with the pairing vertex $\Gamma (k,-k; p, -p)$.  We assume that each of the Fermi surfaces is circular, and consider one of them for definiteness (e.g., the larger Fermi pocket, in which case $k_F \equiv k_{F,L}$).  We follow the standard procedure and expand $
 V^{scr}({\vec k}-{\vec p}) \equiv
 V^{scr}(\theta)$ in the eigenfunctions of the angular momenta in 2D: $\Psi_{l} (\theta) = \sqrt{2} \cos{(l\theta)}$ for $l \neq 0$ and $\Psi_0 (\theta) =1$.  In terms of $\theta$ (the angle between ${\bf k}$ and ${\bf p}$)
\be
 V^{scr}(\theta) = \frac{\pi e^2}{k_F} \frac{\tanh\left[\eta \sin{(\theta/2)}\right]}{\sin{(\theta/2)}}
\label{ch_2}
\ee
The partial components $
 V^{scr}_p (l)$ are
\be
 V^{scr}_p (l) = \frac{\sqrt{2} e^2}{k_F} S_\eta (l)
 \ee
 where
 \be
S_\eta (l) = \int_0^\pi \frac{d\theta}{\sin{\theta}}\  \cos{(2l \theta)}\  \tanh\left[\eta \sin{\theta}\right]
\label{ch_3}
\ee
All $S_\eta (l)$ are positive (repulsive) for any $l$ and $\eta$. At small $\eta$, $S_\eta (l)$ decays exponentially with increasing $l$: $S_\eta (l) \approx 2.6 \eta^{2l+1} e^{-l \times \log {10}}$.

\begin{figure}[tb]
	\centering
	\includegraphics[width=1.1\columnwidth]
{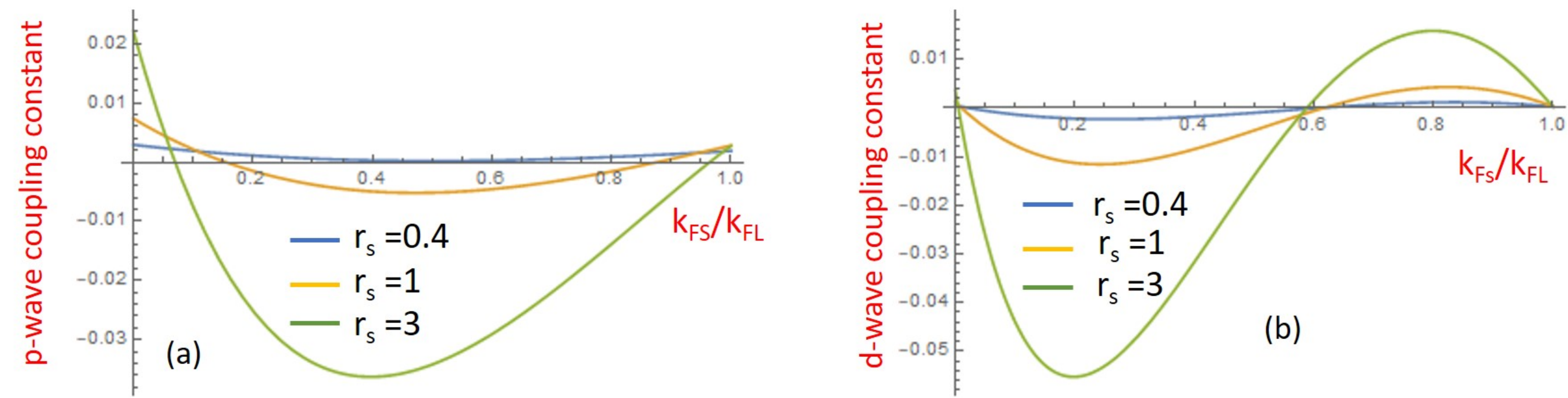}
\caption{Color online. The coupling constants in the $p-$wave  and $d-$wave  channels  (panels (a) and (b)) as functions of the ratio of Fermi momenta $k_{FS}/k_{FL} <1$ for $\alpha=1$ and three different values of $r_s$. Negative value of the coupling means attraction.  The magnitude of the attraction is larger for $p-$wave channel, and the range of $k_{FS}/k_{FL}$, where the attraction holds, is also larger in the $p$-wave channel. }
	\label{fig:1na}
\end{figure}

As already discussed, the fact that the original (bare) interaction is repulsive in all pairing channels is not the end of the story for superconductivity, because the  full pairing interaction is the sum of the bare one and a series of additional terms coming from  the $k$- dependent renormalizations in the particle-hole channel.
 In 2D case, a polarization bubble made out of free fermions with a parabolic dispersion has one-sided non-analyticity:  it is non-analytic at $k =2k_F +0$, but is independent on momentum for all $k < 2k_F$, i.e., for all momentum transfers on the Fermi surface.  As the consequence, the additional interaction that contains a particle-hole bubble has essentially  the same structure of angular components as the bare interaction.
The KL mechanism still applies, but at a higher order in the renormalization, when one includes non-analytic (logarithmic) corrections
 to the particle-hole bubble from vertex renormalization in the particle-particle (Cooper) channel~\cite{third_order}.

 In the case in which there are two Fermi surfaces with $k_{FS}=\eta k_{FL} < k_{FL}$, there is a KL instability\cite{sriandme}
  at the leading (second) order in the coupling, even for rotationally isotropic Fermi pockets in 2D.
  Indeed, consider the pairing interaction between fermions  on a larger Fermi pocket. To second order in the coupling, the renormalization of this interaction  comes from two processes - one involves particle-hole bubble made out of
  electrons from the same larger pocket, another involves particle-hole bubble made out of
  electrons from the smaller pocket. The first renormalization is ineffective, as we just discussed, because particle-hole bubble is independent on momentum for  $q \leq 2k_{FL}$. However, for the second processes, there exists a range of momentum transfers between Fermi momenta on the larger pocket, which exceed $2k_{FS}$. For such momentum transfers, which are in between $2k_{FS}$ and $2k_{FL}$, the particle-hole polarization bubble does depend on momentum, hence it  yields non-zero angular components of the effective interaction
  on the larger Fermi pocket. Below we analyze whether these components are attractive.

\begin{figure}[tb]
	\centering
	\includegraphics[width=1.1\columnwidth]
{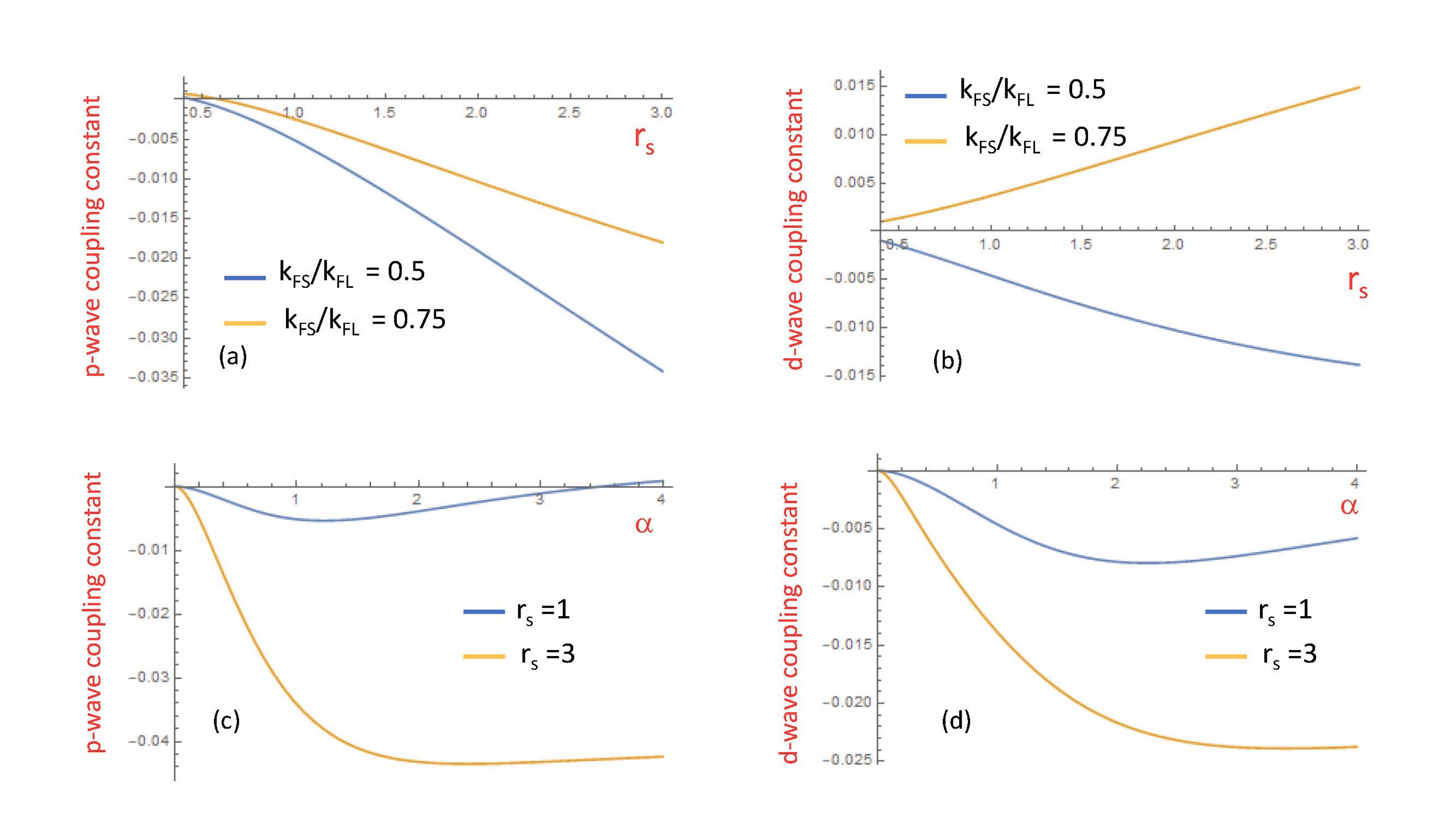}
\caption{Color online. The dimensionless pairing coupling constant $\lambda$ in $p-$wave  and $d-$wave  channels  as functions of $r_s$ for $\alpha=1$ and two different values of  $k_{FS}/k_{FL} =0.5$ and $k_{FS}/k_{FL} =0.75$ (panels (a) and (b)),  and as functions of $\alpha$ for $k_{FS}/k_{FL} =0.5$ and two different $r_s =1$ and $r_s =3$ (panels (c) and (d)).  Observe that the coupling constants in both channels become weakly dependent on $\alpha$ once $r_s$ gets larger, and that the attraction in $d-$wave channel only holds if two Fermi momenta have substantially different values.}
	\label{fig:1nb}
\end{figure}

We  did the analysis for two different forms of the effective pairing interaction, $V^{eff} (q)$, where ${\bf q} = {\bf k} - {\bf p}$ is the momentum transfer between fermions on the larger Fermi pocket.    First, we analyzed $V^{eff} (q)$, taken to second order in  perturbation, i.e.,
\be
 V^{eff}(q) ={
 V}^{scr}(q)\left [ 1- 2{
  V}^{scr}(q)\Pi (q)\right]
 \label{2compVeff}
\ee
where $\Pi (q)$ is the static particle-hole polarization bubble (the density-density susceptibility) made out of fermions near the smaller Fermi surface
(note that in our case with  density-density interaction between fermions from different bands, the  renormalization of $V^{scr}$ into $V^{eff}$   comes only from the bubble diagram).
Second, we analyzed $V^{eff} (q)$, in which we summed up series of bubble renormalizations (
as in RPA). This $V^{eff} (q)$ is
\be
 V^{eff}(q) =\frac{{
 V}^{scr}(q)}{1 + 2 {
 V}^{scr}(q)\Pi (q)}
 \label{2compVeff_1}
\ee
We obtained very similar results in the two cases.  For definiteness, below we present the results for the effective interaction given by
(\ref{2compVeff_1}).

To begin with, we analyzed angular momentum components $V^{eff} (l)$  of the RPA interaction for the case when $k_{FL} = k_{FS}$,
i.e, $\eta = k_{FS}/k_{FL}=1$.
 Then
 $\Pi (q) = \Pi (0) =m/(2\pi)$, and we obtain
\be
 V^{eff}(q) =\frac{\pi e^2}{k_F}
 \left\{ \frac{\tanh\left[ \alpha\sin{(\theta/2)}\right]}{{\sin{(\theta/2)}} + \gamma_{\eta =1} {\tanh\left[ \alpha\sin{(\theta/2)}\right]}} \right\}.
 \label{2compVeff_2}
\ee
 where $\gamma_{\eta =1} \equiv \frac{2\pi e^2 \Pi (0)}{k_F}$.  As expected, all $V^{eff} (l)$  in this case are repulsive and decay exponentially with increasing $l$.

\begin{figure}[tb]
	\centering
	\includegraphics[width=0.8\columnwidth]
{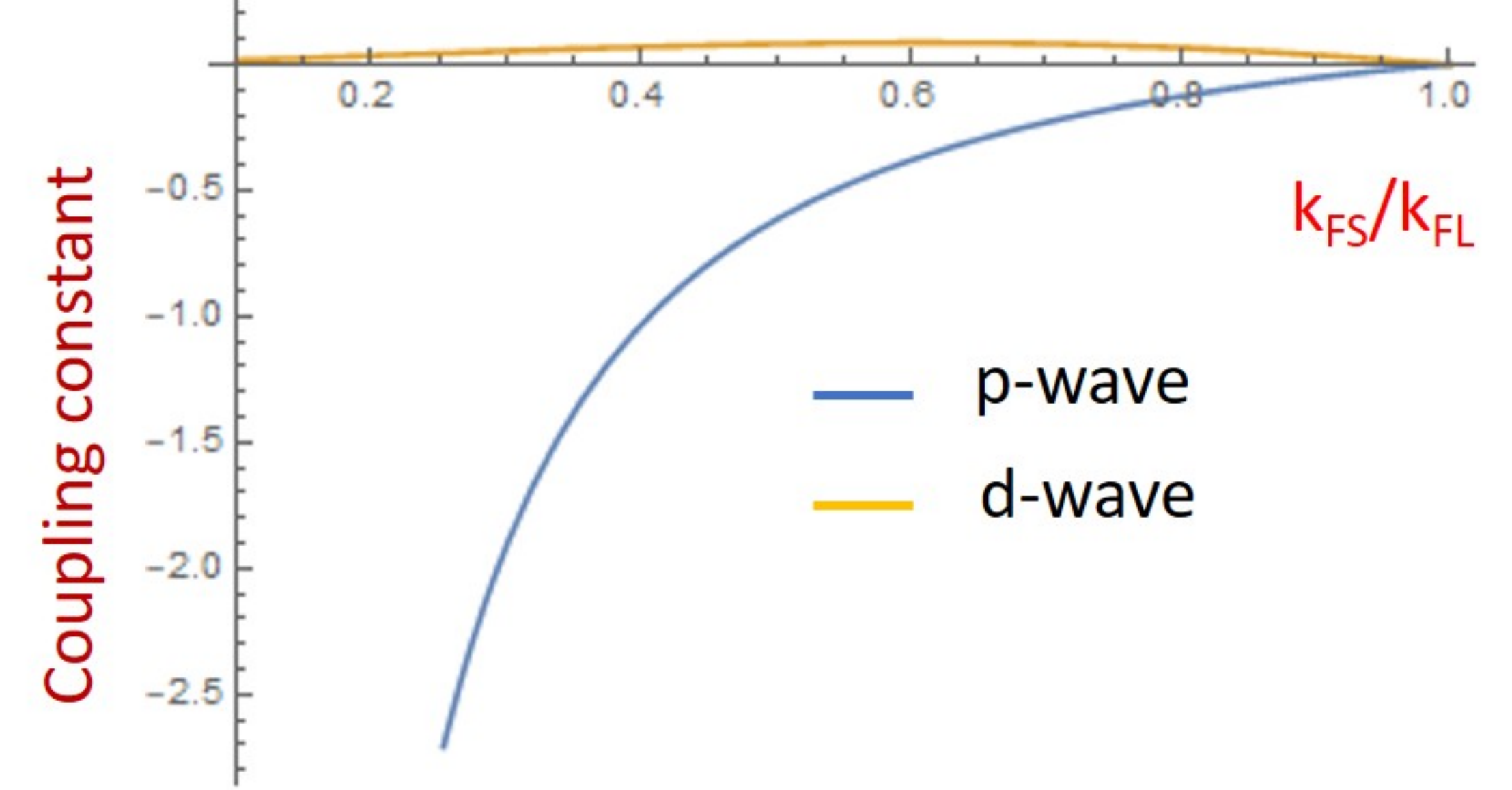}
\caption{Color online: The coupling constants in $p-$wave  and $d-$wave  channels  as functions of $k_{FS}/k_{FL}$ in the formal limit  $r_s \to \infty$.
The coupling in the p-wave channel is attractive, and its magnitude increases as the ratio $k_{FS}/k_{FL}$ decreases.  The coupling in the d-wave channel is repulsive for all values of $k_{FS}/k_{FL}$. When $r_s$ is large but finite, there is an attraction in the d-wave channel, but only for small  $k_{FS}/k_{FL}$.}
	\label{fig:2}
\end{figure}

We now analyze angular momentum  components $V^{eff} (l)$ in our case, when $k_{FL} > k_{FS}$,
 i.e., $\eta <1$.   The polarization bubble $\Pi {q} = \Pi (|{\bf k} - {\bf p}|) = \Pi (\theta)$  is  the Lindhard function
\be
\Pi (\theta) = \frac{m}{2\pi} \left(1 - \theta \left(|\sin{\theta/2}| - \eta \right)\right)  \sqrt{1 - \left(\frac{\eta}{\sin{\theta/2}}\right)^2}.
\ee
 The components $V^{eff} (l)$ are given by
\be
V^{eff} (l) = \frac{\sqrt{2} e^2}{k_F}  S (l),
\ee
where
\begin{widetext}
\ba
S (l) &=&  \int_0^{\arcsin{\eta}} d \theta \ \cos(2l \theta)\left\{\frac{\tanh\left[\alpha \sin{\theta}\right]}
{ \sin(\theta) + \gamma
 \tanh\left[\alpha \sin{\theta}\right]}\right\}
 \nonumber \\
&& + \int_{\arcsin{\eta}}^{\pi/2} d\theta  \cos{(2l \theta)} \left\{\frac{\tanh\left[\alpha \sin{\theta}\right]}
{\sin(\theta) +\gamma
 \left(1 - \sqrt{1 -  \left(\frac{\eta}{\sin{\theta}}\right)^2}\right)\tanh\left[\alpha \sin{\theta}\right]}\right\}
 \label{ch_4}
 \ea
 \end{widetext}
 where $\gamma=  2\pi e^2 \Pi (0)/k_{FL} = m e^2/k_{FL} = r_s/\sqrt{2}$.
  The dimensionless coupling $\lambda (l)$,
 is
 \be
 \lambda (l) = \frac{r_s}{2\pi} S(l)
 \label{lambda}
 \ee

We computed $\lambda(l)$ numerically for different values of $\eta$ and $r_s$.  We found an attraction in $l=1$ and $l=2$ channels (p-wave and d-wave, respectively), in a sizable range of parameters.
The attraction in the $p-$wave channel is stronger.  We show  the results  in Figs. \ref{fig:1na} and \ref{fig:1nb}. In Fig. \ref{fig:1na} we plot coupling constants in $p-$wave and $d-$wave channels (panels (a) and (b)) as functions of the ratio of Fermi momenta $\eta \equiv k_{FS}/k_{FL} <1$ for $\alpha=1$ and three different values of $r_s$. In Fig. \ref{fig:1nb} we plot the same couplings constants in the two channels as functions of $r_s$ for $\alpha=1$ and two different $k_{FS}/k_{FL}$ (panels (a) and (b)) and as functions of $\alpha$ for
$\eta=0.5$ and two different $r_s$ (panels (c) and (d)).    For small $r_s$, $\lambda(l)$ is a linearly increasing function of $r_s$.

The expressions for $\lambda(l) $ simplify (and are $\alpha$ independent) if we formally take the large $\gamma$ (i.e. large $r_s$) limit.
 Here
\begin{widetext}
\be
 \lambda (l) = \frac{1}{\sqrt{2}\pi} \left[ \int_0^{\arcsin{\eta}} d \theta \cos{2l \theta} + \int_{\arcsin{\eta}}^{\pi/2} d\theta  \frac{\cos{2l \theta}}{1 - \sqrt{1 -  \left(\frac{\eta}{\sin{\theta}}\right)^2}}\right]
 \label{new_1}
 \ee
 \end{widetext}
  One can easily check that $\lambda (l=1)$ is negative, of order one, and its magnitude increases
  with decreasing $\eta$. We show $\lambda (l=1)$ and $\lambda (l=2)$ in Fig. \ref{fig:2}.  The KL approach is indeed highly questionable at $r_s \gg 1$
  as there can be additional contributions to the pairing vertex,
   as well as  a variety of  competing instabilities. Still,
   it is encouraging that the  $\lambda (1)$ remains finite in this limit and numerically is not small over a wide range of  $k_{FS}/k_{FL}$.

\section{Conclusions}
\label{conclusions}
In this communication we considered a KL scenario for superconductivity in a two-dimensional electron gas  confined to a narrow well between two metallic planes, for the range of parameters in which there are two occupied subbands with Fermi momenta $k_{FL} > k_{FS}$. The bare interaction between fermions is repulsive, but we found that the effective interaction between fermions on the larger Fermi surface, renormalized by fermions from the  band with smaller $k_{FS}$,  is attractive in non-s-wave %components.
channels.
  The largest attraction is typically in the $p-wave$ channel.
  We analyzed the dependence of the  $p-$wave coupling constant  on the
  system geometry and  found optimal conditions for superconductivity for a given value of the interaction parameter $r_s$.  Overall, the  p-wave coupling increases with $r_s$.

{\bf Acknowledgements}   We thank Sri Raghu, Rui-Rui Du and Loren N. Pfeiffer  for useful discussions.
 This work was supported  by  the  NSF DMR-1523036 ( AVC) and DMR-1608055 (SAK).
  The authors  are thankful to KITP at UCSB,  where part of the work has been done. KITP is supported by NSF grant PHY-1125915.

\end{document}